\documentclass[aps,prl,reprint,superscriptaddress,citeautoscript]{revtex4-1}
\usepackage{graphicx}
\usepackage{natbib}
\usepackage{floatrow}
\usepackage{float}
\usepackage{amsmath}
\usepackage[toc,page]{appendix}
\usepackage{array}
\usepackage{color}
\usepackage{leftidx}
\usepackage[toc,page]{appendix}
\usepackage{xcolor}
\usepackage{caption}
\usepackage{hyperref} 
\hypersetup{
  colorlinks   = true, 
  urlcolor     = black, 
  linkcolor    = blue, 
  citecolor   = blue 
}

\newcommand*\Bell{\ensuremath{\boldsymbol\ell}}
\begin{document}
\def\mean#1{\left< #1 \right>}
\title{Reaction blockading in charged-neutral excited-state chemistry at low collision energy}
\author{Prateek Puri}
\affiliation{Department of Physics and Astronomy, University of California -- Los Angeles, Los Angeles, California, 90095, USA}
\email{prateek.puri01@gmail.com}
\author{Michael Mills}
\affiliation{Department of Physics and Astronomy, University of California -- Los Angeles, Los Angeles, California, 90095, USA}
\author{Ionel Simbotin}
\affiliation{Department of Physics and Astronomy, University of California -- Los Angeles, Los Angeles, California, 90095, USA}
\author{John~A.~Montgomery, Jr.}
\affiliation{Department of Physics, University of Connecticut, Storrs, Connecticut 06269, USA}
\author{Robin C\^ot\'e}
\affiliation{Department of Physics, University of Connecticut, Storrs, Connecticut 06269, USA}
\author{Christian Schneider}
\affiliation{Department of Physics and Astronomy, University of California -- Los Angeles, Los Angeles, California, 90095, USA}
\author{Arthur G. Suits}
\affiliation{Department of Chemistry, University of Missouri, Columbia, MO 65211, USA}
\author{Eric R. Hudson}
\affiliation{Department of Physics and Astronomy, University of California -- Los Angeles, Los Angeles, California, 90095, USA}
\affiliation{UCLA Center for Quantum Science and Engineering, University of California -- Los Angeles, Los Angeles, California, 90095, USA}

\date{\today}

\begin{abstract}

We study an excited atom-polar molecular ion chemical reaction (Ca$^*$ + BaCl$^+$) at low temperature by utilizing a hybrid atom-ion trapping system. The reaction rate and product branching fractions are measured and compared to model calculations as a function of both atomic quantum state and collision energy. At the lowest collision energy we find that the chemical dynamics dramatically differ from capture theory predictions and are primarily dictated by the radiative lifetime of the atomic quantum state instead of the underlying excited-state interaction potential. We provide a simple rule for calculating at what temperature this regime, where the collision complex lifetime is longer than the radiative lifetime of the quantum state, is reached. This effect, which greatly suppresses the reactivity of short-lived excited states, provides a means for directly probing reaction range. It also naturally suppresses unwanted chemical reactions in hybrid trapping experiments, allowing longer molecular ion coherence and interrogation times. 

\end{abstract}

\maketitle

Over the last decade, techniques from ultracold physics have been adapted to the study of chemical systems, bringing unique capabilities including precise control of the reagent quantum states and energy~\cite{Klein2016,Carr2009,Doyle2016,Trippel2013}. While early work focused on all-neutral chemistry, more recently there has been a shift to the study of charged-neutral reactions~\cite{Hall2013,Rellergert2011,Zhang2017, Ratschbacher2012, Sikorsky2018}, as available techniques allow probing a wider range of energy~\cite{Beyer2018} and species, as well as trapping and study of reaction products~\cite{Schneider2016, Schowalter2012,Schmid2017}. Already, these so-called hybrid systems have been used to study reactions of several atom-ion combinations~\cite{Tomza2017,Yang2018,Hall2012,Zipkes2010b,Zipkes2010}, showing a dependence of reactivity on molecular conformation~\cite{Chang2013} and the production of novel molecules~\cite{Puri2017}. Despite this work, there has yet to be a study of atom-polar molecular ion chemistry in these systems. Given that such reactions play a central role in chemistry of the interstellar medium~\cite{Stancil1996,Smith1992,Reddy2010}, which provides the raw materials from which stars, planets, and potentially even life developed, understanding these reactions at low temperature is a fundamental goal for chemistry and physics. Moreover, these same reactions could severely limit experiments aiming to produce quantum-state-selected polar molecular ions~\cite{Calvin2018,Shi2013,Wolf2016} via sympathetic cooling~\cite{Rellergert2012,Hudson2016,Hauser2015} for quantum logic applications~\cite{Hudson2018}. \\
\indent Here, we advance these fronts by using a hybrid trap to study the reaction between electronically-excited Ca atoms and BaCl$^+$ molecules. Using the capabilities of the hybrid trap, we measure the reaction rates and product branching fractions of these reactions at collision energies from $15$ K down to $0.2$ K (all temperatures in this work refer to collision energies in units of J/$k_B=$ K, where $k_B$ is the Boltzmann constant). At the lowest energies in our study, which are amongst the lowest ever studied in a molecular ion-atom system~\cite{Mulin2015,Hauser2015,Allmendinger2016,Hawley1991}, we find a chemical regime where the chemical dynamics are primarily dictated by the radiative lifetime of the reagent quantum state instead of the underlying excited-state interaction potential. Additionally, we provide a simple rule for calculating at what temperature this regime, where the collision time is longer than the radiative lifetime of the quantum state, is reached. \\
\indent This result parallels previous work in excited-state ultracold neutral-neutral systems where reduced reaction rate constants have been observed and explained as a consequence of spontaneous emission suppressing short-range excited-state population~\cite{Julienne1989,Gallagher1989}. Subsequent studies also demonstrated that external optical fields could be used to modify radiative dynamics and directly control reaction outcome~\cite{Weiner1999,Gould1998}.\\
\indent The work presented here extends these techniques to the rapidly developing field of cold molecular-ion chemistry. Specifically, the phenomenon observed here should be universal to atom-ion chemical systems and, through its dependence on the reactive trajectory, provides a means to probe the range of a chemical reaction. It also greatly suppresses the reactivity of short-lived excited states. Therefore, this work implies that care must be taken when interpreting low temperature atom-ion reaction data and that certain unwanted chemical reactions in hybrid trapping experiments can be mitigated by simply going to low temperature and thereby allowing longer molecular ion coherence and interrogation times.\\
\indent In the remainder of this work, we first describe the experimental system and technique for energy control and then present the observed total reaction rates and branching fractions of the Ca $^1$P$_1$ and $^3$P$_2$ states, which show very different behavior. We then describe a qualitative model for the observed effect that provides a simple means to calculate at what temperature this radiative regime is reached. Finally, we compare our experimental results to a more rigorous model of the observed effect that is integrated into a modified long-range capture theory. 

\begin{figure*}
\centering
\includegraphics[width=\textwidth]{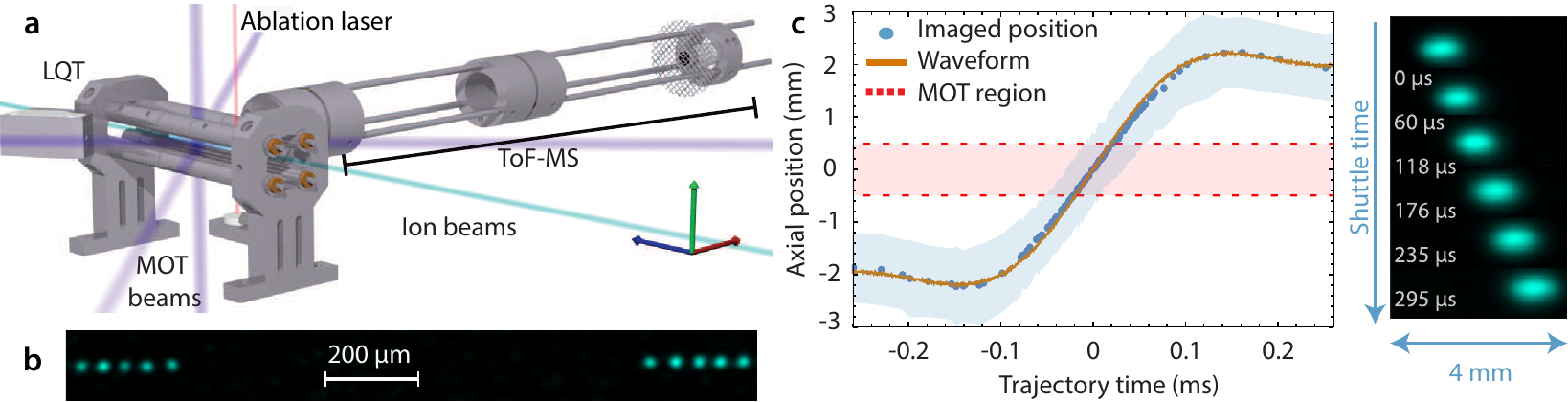}
\caption{\textbf{Experimental apparatus and techniques} \\
\textbf{(a)} The MOTion trap atom-hybrid trap apparatus \textbf{(b)} Image of an ion chain being shuttled over a distance of $\approx 1$~mm at a collision energy of $\approx 750$ mK. To reduce secular heating, the ions spend over $90\%$ of the time at the trajectory endpoints, and thus ion fluorescence is only visible at these locations. \textbf{(c)} The trajectory of a shuttled ion sample, as determined by fluorescence images acquired by triggering on the phase of the shuttling waveform. Also presented is the location of the potential minimum of the axial potential as predicted from the endcap waveform voltages at particular instances of time. For reference, the blue shaded region represents the $1/e$ spatial density width of the three-dimensional Coulomb crystal used in the measurement. Additionally the horizontal red shaded region represents the $1/e$ spatial distance of the MOT cloud. To the right of the plot, an inset displays experimental false-color fluorescence images of the shuttled ions at various times along the shuttling trajectory.}
\label{fig:EXPAPP}
\end{figure*}

\section{Results} 
\label{pathway}
\subsection{Experimental system}

The atom-ion apparatus utilized in this experiment (Fig.~\ref{fig:EXPAPP}a), dubbed the MOTion trap and described elsewhere~\cite{Puri2017, Rellergert2012, Schowalter2016}, is a hybrid system consisting of a co-located magneto-optical trap (MOT) and a linear quadrupole ion trap (LQT) that is radially coupled into a time-of-flight mass spectrometer (ToF)~\cite{Schneider2016, Schowalter2012}. Ba$^+$ ions are co-loaded into the trap and laser-cooled to temperatures of $\approx100$ mK to provide sympathetic cooling for the reactant BaCl$^+$ molecules. To tune reactant collision energy, we employ both a recently developed ion-shuttling technique~\cite{Puri2018} as well as the traditional method of micromotion energy tuning through crystal sample size modulation~\cite{Grier2009,Haze2013}. The former technique, which may be used with both three-dimensional structures and linear ion chains (Fig.~\ref{fig:EXPAPP}b), utilizes precise control of the endcap electrode voltages within the ion trap to modulate the position of the ion at a fixed velocity (Fig.~\ref{fig:EXPAPP}c). This allows for collision energy scanning without problematic effects associated with micromotion, such as micromotion interruption collisions and poor energy resolution~\cite{Chen2014,Rouse2017}. For the reaction rate data discussed below, we implement the technique with three dimensional structures to increase data acquisition throughput. Additionally, while not measured directly, we expect that the internal degrees of freedom of the reactant BaCl$^+$ molecules are cooled via sympathetic cooling collisions~\cite{Rellergert2012} with the Ca MOT. 

\subsection{Observation of reaction blockading}

The reaction is energetically forbidden in the ground state; however, by using previously established methods involving optical pumping and magnetic trapping~\cite{Puri2017}, the reaction is shown to proceed via the Ca $^1$$P_1$ and Ca $^3$$P_J$ electronic states (Appendix Fig.~\ref{fig:RPC}). After identifying the reaction pathways of the system, the dependence of reaction rate on collision energy was explored and is presented in Fig.~\ref{fig:rs}a. All presented theory curves are averaged over the energy distribution of the ions before comparison with the data, and the theory error bands are determined by uncertainties in the polarizability and quadrupole moment values used to construct the molecular potentials utilized in the calculation.  

For the $^3$$P_2$ state, a BaCl$^+$ sample was overlapped with a magnetic trap of pure triplet atoms while micromotion energy tuning was used to change the reactant collision energy from $\approx1-20$ K. The measured reaction rate appears to increase at low energy as expected for an ion-quadrupole reaction and agrees with a long-range capture theory calculation, to be described later. 

Two methods, excess micromotion energy tuning through crystal size manipulation and ion shuttling, were used to measure the collision energy dependence of the Ca ($^1$$P_1) +$ BaCl$^+$ reaction. Over their common range (4 K - 20 K), the two methods agree and reveal an essentially energy-independent reaction rate constant. However, interestingly, unlike the Ca ($^3$$P_2$) state, the measured rate constant \emph{decreases} at low temperature instead of increasing as predicted by standard quadrupole-ion capture theory.

\subsection{Branching fraction analysis}
\label{BRA}

\begin{figure*}
\includegraphics[width=\textwidth]{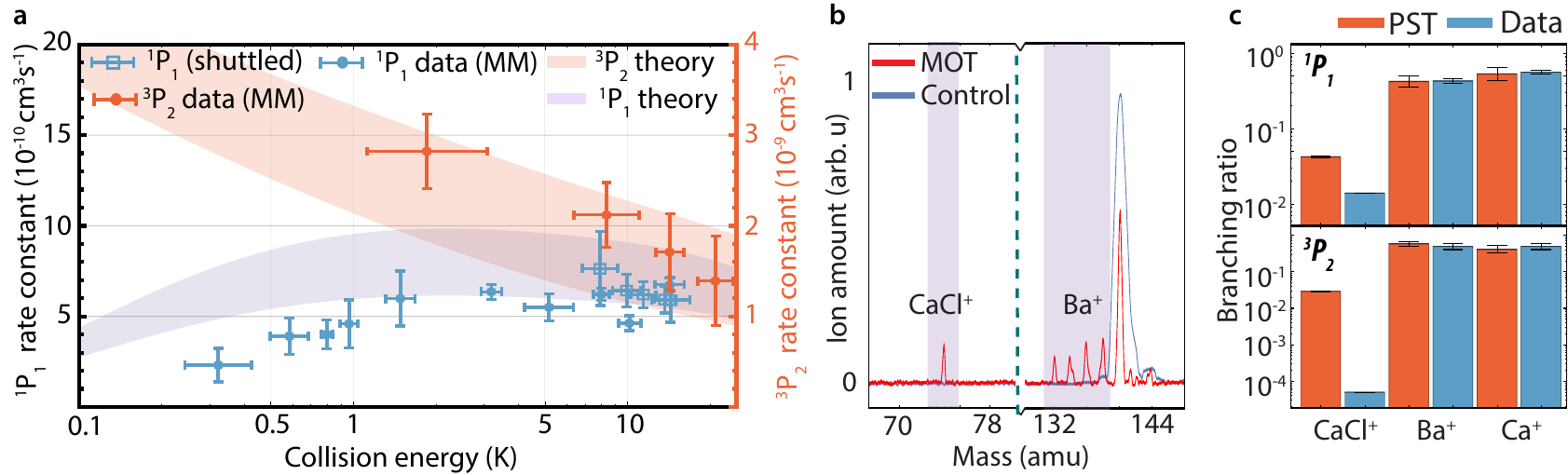}
\caption{\textbf{Reaction blockading in excited neutral-ion systems} \\ 
\textbf{(a)} The experimental dependence of reaction rate constant on collision energy, as measured through both micromotion (MM) tuning (circles) and shuttling (squares) for both the singlet and triplet reaction surfaces. Note that the y-axis scale is different for the two reactions. Both data sets are in reasonable agreement with a modified capture theory incorporating reaction blockading, with the reaction rate of the short-lived Ca $^1$P$_1$ state significantly suppressed at low temperatures as compared to its standard capture theory prediction. For the triplet data, an absolute rate constant is measured at $10$ K and all subsequent data points are normalized with respect to this value due to technical difficulties associated with frequent magnetic trap density measurements. Each data point consists of approximately $100$ measurements, and all errors are expressed at the 1$\sigma$ level. \textbf{(b)} Mass spectra, obtained from the ToF-MS, of the identified product ions of the reaction. The shaded portions identify the masses corresponding to the product ions, and a control spectrum is included where the ions were ejected into the ToF-MS without MOT exposure. \textbf{(c)} A comparison of the measured branching fractions and the predictions of the statistical phase space theory (PST) for both Ca singlet (top) and triplet (bottom) reactions. Errors are expressed at the $1\sigma$ level and, in the case of the CaCl$^+$ values, may be smaller than the plot-marker size. }
\label{fig:rs}
\end{figure*}

Given this departure from standard capture theory, we then measured the product branching fractions to gain a fuller understanding of the chemical dynamics. For experimental convenience, reactions between Ca and non-shuttled BaCl$^+$ ions were studied at an average energy of $\approx 5$~K. For reactions with the Ca $^1$P$_1$ and Ca $^3$P$_2$ states, there are three energetically allowed pathways: 
	
\begin{align} \label{eqn:CaClp}
\mathrm{BaCl^+ + Ca } &\mathrm{ \longrightarrow CaCl^+ + Ba} \\
&\mathrm{\longrightarrow  Ba^+ + CaCl} \\
&\mathrm{\longrightarrow  Ca^+ + BaCl} \label{eqn:CEX}
\end{align}

Products from the first two reactions are experimentally identified by the appearance of reaction products in ToF spectra (see Fig.~\ref{fig:rs}b). While the reaction products are created with $\lesssim$ $1$ eV of energy, the radial (axial) trap depth of the LQT is $\approx 4 (0.5)$ eV, and thus, $\geq~95\%$ of charged products are expected to be recaptured in the LQT, assuming isotropic scattering of ions after reaction. Products from the direct charge exchange reaction (equation (\ref{eqn:CEX})) cannot be inferred directly due to a background Ca$^+$ influx from MOT atom photoionization; thus, these products are inferred indirectly through the presence of the other two products. Additionally, Ba$^+$ products are distinguished via mass from the isotopically-pure $^{138}$Ba$^+$ coolant ions since the initial BaCl$^+$ reactant sample is present in natural abundance.

By monitoring the appearance of Ba$^+$ and CaCl$^+$ ions in the ToF-MS spectra, branching fractions, $\gamma_{i}$, defined as the number of product ions in the $i^{th}$ exit channel formed per BaCl$^+$ loss event, are measured. For the $4s4p$ $^{1}P_1$ entrance channel, $[\gamma_{\rm{CaCl^+}},\;\gamma_{\rm{Ba^+}},\;\gamma_{\rm{Ca^+}}] = [0.014(4),\;0.43(6),\;0.57(6)]$, while $[\gamma_{\rm{CaCl^+}},\;\gamma_{\rm{Ba^+}},\;\gamma_{\rm{Ca^+}}] = [0.0001(8),\;0.5(2),\;0.5(2)]$ is measured for the $4s4p$ $^{3}P_2$ state. Notably, the CaCl$^+$ molecule is only definitively detected in $^1$$P_1$ reactions, providing a means for quantum state control of reaction products. 

Also shown in Fig.~\ref{fig:rs}c are the branching fractions predicted by a phase space theory calculation~\cite{Pechukas1966}. This calculation assumes all product states that are accessible via energy and angular momentum conservation are equally probable. Thus, product branching fractions are calculated by counting the total number of states available to each reaction product (Appendix~\ref{phasespace}). The model predicts branching fractions of $[\gamma_{\rm{CaCl^+}},\;\gamma_{\rm{Ba^+}},\;\gamma_{\rm{Ca^+}}] = [0.04(2),\;0.42(17),\;0.53(19)]$ for the singlet and $[\gamma_{\rm{CaCl^+}},\;\gamma_{\rm{Ba^+}},\;\gamma_{\rm{Ca^+}}] = [0.018(14),\;0.56(21),\;0.41(23)]$ for the triplet, which are in reasonable agreement with the experimental values (Fig.~\ref{fig:rs}c). The error bars are primarily determined by energetic uncertainties in the exothermicity of each reaction channel. The relative ordering of the branching fractions can be attributed to two main factors. First, the CaCl$^+$ exit channel has the lowest product exothermicity and therefore the fewest accessible rovibronic states. Second, the ground state of the CaCl$^+$ + Ba asymptote is composed of two singlets, whereas the other asymptotes are composed of two doublets, reducing the number of states accessible to CaCl$^+$ by approximately a factor of four. \\
\indent The relatively good agreement of this model with the data suggests that the reaction proceeds via a long-lived collision complex, which facilitates the realization of ergodicity and therefore the statistical assumption of the model. 

\subsection{Modeling of reaction blockading} 
\label{radsupp} 

Given the evidence for a long-lived collision complex from the product branching data and the dramatic difference in reactivity as a function of temperature for quantum states with a long ($^3$P$_2$, $\tau \approx 118$ min) and short ($^1$P$_1$, $\tau \approx 4$ ns) radiative lifetime, the observations suggest that that spontaneous emission modifies the chemical dynamics. Because any reaction on an excited surface starts in the separated atom limit, the reagents must propagate inward to short-range separation ($\sim10$ $a_0$) before a chemical reaction can occur. If the time it takes to propagate inward to form a collision complex and pass through the transition state to products is similar to the radiative lifetime of the excited reagent, it is likely that spontaneous emission will occur during the chemical event. In this limit, which is more likely at extremely low temperatures, the reactivity of excited reagents will be given by the reactivity of the surface reached through spontaneous emission, which in the present case is the endoergic ground-state surface.  

To estimate the temperature at which this effect becomes important, it is necessary to calculate the dependence of the total collision time on temperature. Normally, this would be estimated by calculating the lifetime of the three-body collision complex from Rice-Ramsperger-Kassel-Marcus theory~\cite{Rice1927,Dagdigian1977}; however, this lifetime, which is typically a few vibrational periods, severely underestimates the total collision time at low temperature as it neglects the time it takes for the reagents to fall into the collision complex.

Following similar approaches in neutral-neutral systems~\cite{Weiner1999}, to account for this effect we consider the collision trajectory of the reactants as they spiral inward along their ground-state and excited-state surfaces. Since these surfaces have different long-range forms, due to their differing polarizabilities and quadrupole moments, they diverge from one another as the atom-ion separation distance decreases. This causes any lasers resonant with the system in the dissociation limit to become far-detuned in the region where chemical dynamics occur. At cold temperatures, this effect, dubbed reaction blockading, makes it unlikely that the Ca atom will remain in the excited state long enough to react before spontaneously emitting; however, at higher collision velocities, such events are less likely to occur before the atom reaches short-range. 

This effect is particularly sensitive to the atomic lifetime of the reactive state, as longer excited-state lifetimes allow the reaction complex to reach short-range more easily before being interrupted by a spontaneous emission event. As a result, once reaction blockading is integrated into the capture theory predictions, we observe good agreement with the Ca $^1$P$_1$ data; whereas, for the long-lived triplet state, the effect of reaction blockading is negligible, as expected (Fig.~\ref{fig:rs}a).    

While a more quantitatively rigorous model of the blockading effect is presented in the next section, we first develop a simple model to estimate when radiative effects become significant. The collision energy, $E_{B}$, at which the blockading effect reduces the total reaction rate by 1/2 can be approximated by considering the amount of time, $t_{B}=\tau \ln(2)$, it takes to deplete the excited state population by the same amount, where $\tau$ is the excited-state lifetime. 

From conservation of energy, $E_{tot} = E_{col}(r) + V_{ex}(r)$, where $E_{tot}$ is the total energy of the system and $V_{ex}(r)$ is the excited-state potential of the system, and thus, $t_{B}$ can be expressed in terms of the collision energy as
\begin{equation} \label{complexRS}
\tau \ln(2) = \int_{r_c}^{r_s}\left(\frac{\mu}{2[E_{tot}-V_{ex}(r)]}\right)^{1/2} dr 
\end{equation}
where $r_s$ is the short-range distance at which the chemical event occurs ($\approx50$ $a_0$ in our system) and $r_c$ is the critical internuclear separation distance where the addressing laser becomes detuned from its associated atomic transition ($\approx1200$ $a_0$ in our system). We obtain $r_c$ by solving $V_{ex}(r_c)=V_{gs}(r_c)+\Delta E-\hbar \Gamma$ where $V_{ex}(r)$ and $V_{gs}(r)$ are the excited-state and ground-state molecular potentials of the system, $\Delta E$ is the separation-limit energetic difference between the ground and excited states, and $\Gamma$ is the linewidth of the neutral cooling transition. 

Further, if we approximate the velocity, and thus the kinetic energy, of the system as being constant during the latter portion of the trajectory, then $E_{tot}-V_{ex}(r)\approx E_{B}$ and equation~(\ref{complexRS}) can be inverted to yield the final result
\begin{equation} 
\label{simpleRS}
\begin{split}
E_{B} &= \frac{1}{\ln(2)^2}\frac{(r_c-r_s)^2}{\tau^2} \frac{\mu}{2} \\
&\approx \frac{1}{\ln(2)^2}\frac{r_c^2}{\tau^2} \frac{\mu}{2}
\end{split}
\end{equation}
where $r_s \ll r_c$ in the latter approximation. Applying equation~(\ref{simpleRS}) to the Ca$^*$+BaCl$^+$ system, we calculate $E_{B}$ to be $560$ mK for the $^1$P$_1$ state and $\ll~1~\mu$K for the $^3$P$_2$ state.

\subsection{Modified capture theory} 
\label{MCT}
\begin{figure*}
\centering
\includegraphics[width=\textwidth]{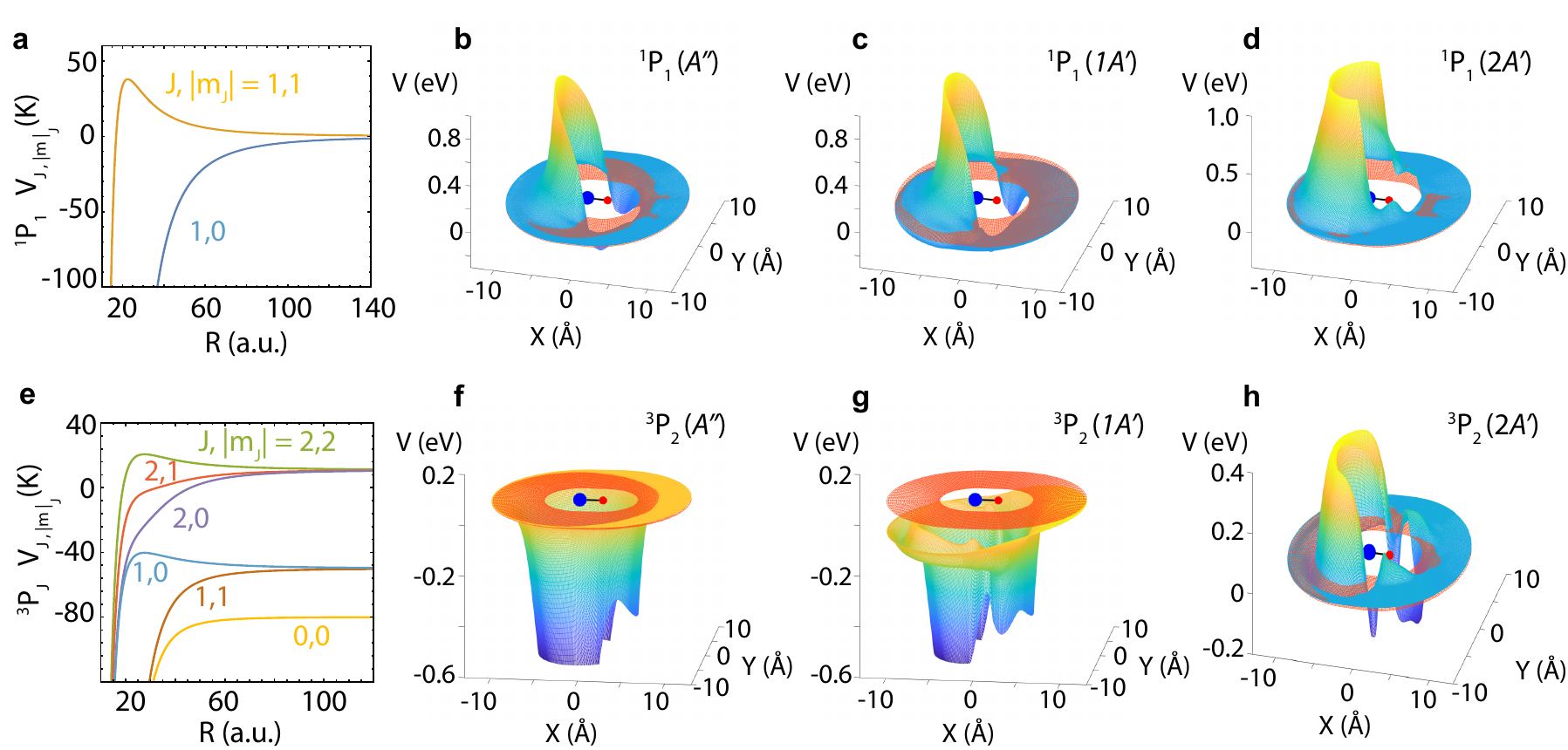}
\vspace{-0mm}
\caption{ \textbf{Potential energy curves and surfaces}. \\ 
\textbf{(a)} Potential energy curves for each (J,$|m_J|$) level expressed as a function of internuclear separation coordinates for BaCl$^+$ + Ca ($^1$P$_1$), where the molecular ion is considered as a point charge placed at the origin (Ba$^+$ in blue, Cl in red) . \textbf{(b-d)} Potential energy surfaces corresponding to \textbf{(b)} the A$^{''}$ symmetry and (\textbf{c} and \textbf{d}) the two $A'$ symmetries for the BaCl$^+$ + Ca ($^1$P$_1$) complex. The $x$ and $y$ axes are in \AA~and the $z$ axis in eV. The orange plane is the asymptotic value 3.08 eV above the global ground state of BaCl$^+$ + Ca, computed with the same level of theory at $R=50$~\AA. Short-range energetic barriers along the $A^{''}$ (b) and $2A^{'}$ (d) surfaces prevent the reactants in these surfaces from reaching the reaction region at short-range, resulting in a reduction of the overall Ca $^1$P$_1$ reaction rate by a factor of $1/3$ (see equation~(\ref{eqn:LReqn})). \textbf{(e-h)} Similarly for the BaCl$^+$ + Ca($^3$P$_2$) complex, potential energy curves \textbf{(e)} and surfaces are displayed corresponding to the $A''$ (\textbf{f}) and the two $A'$ symmetries (\textbf{g} and \textbf{h}), with axes consistent with those of the singlet. The orange plane is the asymptotic value 1.88 eV above the global ground state of BaCl$^+$ + Ca (singlet) computed with the same level of theory at $R=30$~\AA. Unlike the Ca $^1$P$_1$ surfaces, no short-range energetic barriers prevent reaching the reaction region for the triplet surfaces and thus there is no additional reduction in triplet reaction rate.}
\label{fig:PES}
\end{figure*}

For quantitative comparison to the measured rate constant, we require a more rigorous model of the blockading effect than the one presented in the previous section. To this end, we modify standard long-range capture theory to include a probability of reaching short-range before spontaneous emission for each partial wave, $P_{\ell}(E_{col},m_J)$:

\begin{equation} \label{eqn:LReqn}
\begin{split}
\sigma(&E_{col}) \approx\\
 &\frac{\pi \hbar^2}{2 \mu E_{col}} \sum_{m_J} \frac{ \eta_{m_J} \chi_S}{2J+1} \sum_{\ell=0}^{\ell_{\rm{max}}(E_{col},m_J)} (2\ell+1) P_{\ell} (E_{col},m_J) 
\end{split}
\end{equation}
where $\eta_{m_J}$ is the probability that the $m_J$ Zeeman level reacts if it reaches short-range, $\chi_S$ is the probability that the reaction does not produce an excited BaCl$^+$ molecule, and $\ell_{\rm{max}}(E_{col},m_J)$ is the maximum orbital angular momentum that the entrance system can possess at energy $E_{col}$ while still being able to reach short-range. $P_\ell (E_{col},m_J)$ is calculated by simultaneously finding the trajectory of colliding pair on their interaction potential and solving a two-level Einstein rate equation model to account for electronic population dynamics in the presence of changing electronic energy levels (Appendix~\ref{MCT}).

The long-range potentials, assuming the ion to be a positive point charge, are dominated by the ion-dipole polarizability and ion-quadrupole terms as shown in Fig.~\ref{fig:PES}a,d. As can be seen, for the  $^3$$P_2$ and $^1$$P_1$ levels the $(J,|m_J|)$ =(2,2) and (1,0), and $(J,|m_J|)$=(1,1)  states, respectively, have large barriers. Reactants on these potentials cannot reach short-range on any partial wave. While the other potentials are attractive at long-range, it is necessary to match the long-range potential to the short-range potential energy surface for a complete reaction rate calculation. 

To this end, electronic structure calculations were performed~\cite{Frisch2009,Werner2011} to calculate the reaction surface for each excited-state reagent, using equation-of-motion coupled cluster theory including single and double excitations (EOM-CCSD) (Appendix~\ref{theory}). The resulting potential surfaces for separation between BaCl$^+$ and Ca ranging from $4 \AA$ and $10 \AA$ are shown in Fig.~\ref{fig:PES} for the three excited singlet and triplet symmetries, $1A'$, $2A'$, and $A''$. The dark plane in each panel indicates the asymptotic energy of BaCl$^+$ + Ca -- $3.08$ eV for the singlet and $1.88$ eV for the triplet, neglecting spin-orbit couplings. Most notably, in the case of the singlet, the surfaces corresponding to the $2A'$  and the $A''$ geometry prevent the reactants from approaching the reaction region at short-range. Since only the surface corresponding to the other $A'$ geometry allows approach to the reaction region, $\eta_{m_J}$=$1/3$ for reactions originating in the Ca($^1$$P_1$) state. In the case of the triplet, all three surfaces possess a pathway to reach the reaction region. Therefore, given the low temperatures probed here, we take $\eta_{m_J}$=$1$ for all reactions originating in the Ca($^3$$P_2$) state.

Finally, for the present system, there are many inelastic channels that lead to a loss of the initial reagent population, but result in excited states of [BaCl$^+$ + Ca]  that ultimately radiatively decay back into the [BaCl$^+$ + Ca] ground state, making such reactions indistinguishable from non-reaction events. To estimate the probability of such events, we apply the phase space theory model described earlier to the exit channel product [BaCl$^+$ + Ca]. After including all energetically accessible excited states, we obtain $\chi_S$ = 0.76(13) and $0.72(17)$ for the singlet and triplet channels, respectively, with the errors again determined by uncertainties in exit channel exothermicities (Appendix~\ref{phasespace}). 

The results of this modified long-range capture model, after thermal averaging, are shown in Fig.~\ref{fig:rs}a and are in reasonable agreement with the data. 

\section{Discussion}

The observed reaction blockading is expected to be a general effect in low temperature ion-neutral chemistry, as the monopole field of the ion significantly alters the electronic structure of the neutral at relatively long-range. While the modified capture theory developed here can quantitatively treat the suppression effect, the simple expression presented in equation~(\ref{simpleRS}) can be used to estimate if reaction blockading will be important for a given system. For example, in the (Rb + N$_2^+$)~\cite{Hall2012} and (Rb + Ba$^+$)~\cite{Hall2013} systems studied by the Basel group, equation~(\ref{simpleRS}) predicts that reaction blockading is important at $E\lesssim 10$ mK, significantly below the temperatures of their studies.

In addition to the role reaction blockading may play in the interpretation of low-temperature excited-state reactions, it has an important consequence for the field of quantum-state-selected molecular ions. The reactions studied here represent the dominant loss mechanism for preparing cold molecular ions with laser-cooled neutral atoms~\cite{Hudson2016}, and the described suppression effect may be critical for enhancing sample overlap times in next-generation hybrid trapping experiments~\cite{Staanum2010,Hudson2018}.

In summary, we have presented an investigation of polar molecular ion-neutral chemistry at cold temperature. A recently developed ion-shuttling technique, along with micromotion-energy tuning, was employed to measure the dependence of the reaction rate on collision energy. Branching fractions for the reaction were measured and advanced electronic structure calculations, complemented by a long-range capture theory analysis, were performed to understand the collision dynamics of the system, revealing a strong dependence of the reaction on approach angle of the incoming Ca atom. Further, we have demonstrated that spontaneous emission during the collision strongly affects the reaction rate of the system, resulting in a reaction blockading phenomenon. This effect is incorporated into a modified-capture theory model and compared to the experimental data, demonstrating reasonable agreement. Further, a rule of thumb is developed to estimate at what temperature the reaction blockading effect becomes important for a given chemical system. This work builds on previous studies exploring radiative effects in neutral-neutral reactions~\cite{Weiner1999,Gould1998,Gallagher1989} and represents an important step towards understanding quantum chemical dynamics in hybrid systems and well as controlling such dynamics with optical and electromagnetic fields.

\section{Methods}

\subsection{Experimental apparatus} 
\label{experiment}

In a typical experimental sequence, a sample of Ba$^+$ and BaCl$^+$ ions is loaded into the ion trap by ablating a BaCl$_2$ target with a pulsed $1064$ nm Nd:YAG laser. The Ba$^+$ ions act as a translational sympathetic coolant~\cite{Staanum2010,Rugango2015,Rellergert2012} for the BaCl$^+$ molecules, enabling the study of reactions at low collision energy. After initialization, the ion crystal is immersed, either while being shuttled or while stationary, in a cloud of roughly two million Ca atoms loaded into a MOT. The ions and atoms are allowed to interact for a variable amount of immersion time before the ions are ejected into the ToF, which registers the amount of each species present in the LQT at a given instant. 

\subsection{Reaction rate extraction} 

BaCl$^+$ depletion, caused by reactive collisions with the ultracold Ca sample, is monitored as a function of reaction time, and the reaction kinetics data is fit to a simple rate equation model in order to determine total reaction rate
\begin{equation} \label{rateconstant}
\Gamma_{T} = \sum\limits_{i}{\left\langle\int_{}^{} \hat\rho_{ion}(\vec{r}(t)) \eta_{i}(\vec{r}) k_{i}(E(\vec{r}(t))) d^{3}r\right\rangle}
\end{equation}
where $\Gamma_T$ is the total reaction rate, $\eta_{i}(\vec{r})$ is the atom density of the $i^{th}$ Ca electronic state at position $\vec{r}$, $\hat\rho_{ion}(\vec{r}(t))$ is the integral-normalized spatial ion density an ion sample undergoing a trajectory $\vec{r}(t)$ (time independent if non-shuttled), and $k_{i} (E(\vec{r}(t)))$ is the reaction rate constant of the $i^{th}$ electronic state at the spatially dependent collision energy, $E(\vec{r}(t))$, and the angled brackets represent a time average over the reaction time. Fluorescence from the laser-cooled Ba$^+$ ions and Ca atoms can be imaged with three separate EMCCD cameras, allowing for accurate determination of the spatial distributions and densities needed for the rate constant extraction. 

\subsection{Collision energy control} 

We employ a recently developed~\cite{Puri2018} ion-shuttling technique to measure how the reaction rate varies with reactant collision energy. Collision energy in hybrid trap experiments is typically controlled by tuning the excess micromotion energy of an ion sample within a LQT; however, while this method allows precise control of the average collision energy, it suffers from poor energy resolution and is susceptible to problematic effects such as micromotion interruption heating. The shuttling method improves upon this technique by offering comparatively higher energy resolution while maintaining the ions on the trap null, and the essentials of the method are briefly reviewed below. 

For small displacements around the ion trap center, the potential in the axial dimension can be approximated as a harmonic oscillator, where offsets in voltages applied to opposing endcap electrodes serve to shift the equilibrium position of the harmonic potential. A time-dependent voltage offset between the endcaps provides precise control the velocity of the ion, and thus the collision energy, as a function of time. Additionally, the endcap voltage waveforms must be selected carefully to ensure that the ions respond adiabatically to the translating axial potential and experience minimal secular heating. 

The position of the ions while shuttling can be tracked for ions with velocities $\lesssim50$ m/s by phase triggering an EMCCD camera on the shuttling waveform and obtaining fluorescence images at various positions along the shuttling trajectory, as shown in Fig.~\ref{fig:EXPAPP}b. In this range, the ions follow the waveform predicted trajectory and experience roughly linear motion in the intersection region with the MOT. Further, control measurements are taken by monitoring the ion amount in the LQT while shuttling for several hundred seconds with no MOT present and demonstrate that ion loss due to secular heating is negligible. 

\section{Acknowledgments}

This work was supported by National Science Foundation (PHY-1205311, PHY-1806653, and DGE-1650604) and Army Research Office (W911NF-15-1-0121, W911NF-14-1-0378, and W911NF-13-1-0213) grants. \\ \\

\bibliographystyle{naturemag_nourl}

\section{References}

\begin{thebibliography}{10}
\expandafter\ifx\csname url\endcsname\relax
  \def\url#1{\texttt{#1}}\fi
\expandafter\ifx\csname urlprefix\endcsname\relax\def\urlprefix{URL }\fi
\providecommand{\bibinfo}[2]{#2}
\providecommand{\eprint}[2][]{\url{#2}}

\bibitem{Klein2016}
\bibinfo{author}{Klein, A.} \emph{et~al.}
\newblock \bibinfo{title}{{Directly probing anisotropy in atom–molecule
  collisions through quantum scattering resonances}}.
\newblock \emph{\bibinfo{journal}{Nat. Phys.}} \textbf{\bibinfo{volume}{13}},
  \bibinfo{pages}{35} (\bibinfo{year}{2016}).

\bibitem{Carr2009}
\bibinfo{author}{Carr, L.~D.}, \bibinfo{author}{DeMille, D.},
  \bibinfo{author}{Krems, R.~V.} \& \bibinfo{author}{Ye, J.}
\newblock \bibinfo{title}{Cold and ultracold molecules: science, technology and
  applications}.
\newblock \emph{\bibinfo{journal}{N. J. Phys.}} \textbf{\bibinfo{volume}{11}},
  \bibinfo{pages}{055049} (\bibinfo{year}{2009}).

\bibitem{Doyle2016}
\bibinfo{author}{Doyle, J.~M.}, \bibinfo{author}{Bretislav, F.} \&
  \bibinfo{author}{Edvardas, N.}
\newblock \bibinfo{title}{Physics and chemistry with cold molecules}.
\newblock \emph{\bibinfo{journal}{ChemPhysChem}} \textbf{\bibinfo{volume}{17}},
  \bibinfo{pages}{3581--3582}.

\bibitem{Trippel2013}
\bibinfo{author}{Trippel, S.}, \bibinfo{author}{Stei, M.},
  \bibinfo{author}{Cox, J.~A.} \& \bibinfo{author}{Wester, R.}
\newblock \bibinfo{title}{Differential scattering cross-sections for the
  different product vibrational states in the ion-molecule reaction
  $\mathrm{Ar}^+$ + $\mathrm{N}_2$}.
\newblock \emph{\bibinfo{journal}{Phys. Rev. Lett.}}
  \textbf{\bibinfo{volume}{110}}, \bibinfo{pages}{163201}
  (\bibinfo{year}{2013}).

\bibitem{Hall2013}
\bibinfo{author}{Hall, F.~H.}, \bibinfo{author}{Aymar, M.},
  \bibinfo{author}{Raoult, M.}, \bibinfo{author}{Dulieu, O.} \&
  \bibinfo{author}{Willitsch, S.}
\newblock \bibinfo{title}{Light-assisted cold chemical reactions of barium ions
  with rubidium atoms}.
\newblock \emph{\bibinfo{journal}{Mol. Phys.}} \textbf{\bibinfo{volume}{111}},
  \bibinfo{pages}{1683--1690} (\bibinfo{year}{2013}).

\bibitem{Rellergert2011}
\bibinfo{author}{Rellergert, W.~G.} \emph{et~al.}
\newblock \bibinfo{title}{Measurement of a large chemical reaction rate between
  ultracold closed-shell $^{40}\mathrm{Ca}$ atoms and open-shell
  $^{174}\mathrm{Yb}^{+}$ ions held in a hybrid atom-ion trap}.
\newblock \emph{\bibinfo{journal}{Phys. Rev. Lett.}}
  \textbf{\bibinfo{volume}{107}}, \bibinfo{pages}{243201}
  (\bibinfo{year}{2011}).

\bibitem{Zhang2017}
\bibinfo{author}{Zhang, D.} \& \bibinfo{author}{Willitsch, S.}
\newblock In \emph{\bibinfo{booktitle}{Cold Chemistry: Molecular Scattering and
  Reactivity Near Absolute Zero Ch.10}} (\bibinfo{publisher}{The Royal Society
  of Chemistry}, \bibinfo{address}{London}, \bibinfo{year}{2018}).

\bibitem{Ratschbacher2012}
\bibinfo{author}{Ratschbacher, L.}, \bibinfo{author}{Zipkes, C.},
  \bibinfo{author}{Sias, C.} \& \bibinfo{author}{K{\"{o}}hl, M.}
\newblock \bibinfo{title}{{Controlling chemical reactions of a single
  particle}}.
\newblock \emph{\bibinfo{journal}{Nat. Phys.}} \textbf{\bibinfo{volume}{8}},
  \bibinfo{pages}{649} (\bibinfo{year}{2012}).

\bibitem{Sikorsky2018}
\bibinfo{author}{Sikorsky, T.}, \bibinfo{author}{Meir, Z.},
  \bibinfo{author}{Ben-shlomi, R.}, \bibinfo{author}{Akerman, N.} \&
  \bibinfo{author}{Ozeri, R.}
\newblock \bibinfo{title}{{Spin-controlled atom--ion chemistry}}.
\newblock \emph{\bibinfo{journal}{Nat. Comm.}} \textbf{\bibinfo{volume}{9}},
  \bibinfo{pages}{920} (\bibinfo{year}{2018}).

\bibitem{Beyer2018}
\bibinfo{author}{Beyer, M.} \& \bibinfo{author}{Merkt, F.}
\newblock \bibinfo{title}{Half-collision approach to cold chemistry: Shape
  resonances, elastic scattering, and radiative association in the
  $\mathrm{H}^{+}+\mathrm{H}$ and $\mathrm{D}^{+}+\mathrm{D}$ collision
  systems}.
\newblock \emph{\bibinfo{journal}{Phys. Rev. X}} \textbf{\bibinfo{volume}{8}},
  \bibinfo{pages}{031085} (\bibinfo{year}{2018}).

\bibitem{Schneider2016}
\bibinfo{author}{Schneider, C.}, \bibinfo{author}{Schowalter, S.~J.},
  \bibinfo{author}{Yu, P.} \& \bibinfo{author}{Hudson, E.~R.}
\newblock \bibinfo{title}{Electronics of an ion trap with integrated
  time-of-flight mass spectrometer}.
\newblock \emph{\bibinfo{journal}{Int. J. Mass Spectrom.}}
  \textbf{\bibinfo{volume}{394}}, \bibinfo{pages}{1 -- 8}
  (\bibinfo{year}{2016}).

\bibitem{Schowalter2012}
\bibinfo{author}{Schowalter, S.~J.}, \bibinfo{author}{Chen, K.},
  \bibinfo{author}{Rellergert, W.~G.}, \bibinfo{author}{Sullivan, S.~T.} \&
  \bibinfo{author}{Hudson, E.~R.}
\newblock \bibinfo{title}{An integrated ion trap and time-of-flight mass
  spectrometer for chemical and photo- reaction dynamics studies}.
\newblock \emph{\bibinfo{journal}{Rev. Sci. Instrum.}}
  \textbf{\bibinfo{volume}{83}}, \bibinfo{pages}{043103}
  (\bibinfo{year}{2012}).

\bibitem{Schmid2017}
\bibinfo{author}{Schmid, P.~C.}, \bibinfo{author}{Greenberg, J.},
  \bibinfo{author}{Miller, M.~I.}, \bibinfo{author}{Loeffler, K.} \&
  \bibinfo{author}{Lewandowski, H.~J.}
\newblock \bibinfo{title}{An ion trap time-of-flight mass spectrometer with
  high mass resolution for cold trapped ion experiments}.
\newblock \emph{\bibinfo{journal}{Rev. Sci. Instrum.}}
  \textbf{\bibinfo{volume}{88}}, \bibinfo{pages}{123107}
  (\bibinfo{year}{2017}).

\bibitem{Tomza2017}
\bibinfo{author}{Tomza, M.} \emph{et~al.}
\newblock \bibinfo{title}{{Cold ion-atom systems}} (\bibinfo{year}{2017}).
\newblock \bibinfo{note}{{Preprint at https://arxiv.org/abs/1708.07832}}.

\bibitem{Yang2018}
\bibinfo{author}{Yang, T.} \emph{et~al.}
\newblock \bibinfo{title}{{Optical control of reactions between water and
  laser-cooled Be$^+$ ions}}.
\newblock \emph{\bibinfo{journal}{J. Phys. Chem. Lett.}}
  \textbf{\bibinfo{volume}{9}}, \bibinfo{pages}{3555--3560}
  (\bibinfo{year}{2018}).

\bibitem{Hall2012}
\bibinfo{author}{Hall, F. H.~J.} \& \bibinfo{author}{Willitsch, S.}
\newblock \bibinfo{title}{Millikelvin reactive collisions between
  sympathetically cooled molecular ions and laser-cooled atoms in an ion-atom
  hybrid trap}.
\newblock \emph{\bibinfo{journal}{Phys. Rev. Lett.}}
  \textbf{\bibinfo{volume}{109}}, \bibinfo{pages}{233202}
  (\bibinfo{year}{2012}).

\bibitem{Zipkes2010b}
\bibinfo{author}{Zipkes, C.}, \bibinfo{author}{Palzer, S.},
  \bibinfo{author}{Sias, C.} \& \bibinfo{author}{K$\mathrm{\ddot{o}}$hl, M.}
\newblock \bibinfo{title}{A trapped single ion inside a bose-einstein
  condensate}.
\newblock \emph{\bibinfo{journal}{Nature}} \textbf{\bibinfo{volume}{464}},
  \bibinfo{pages}{388--391} (\bibinfo{year}{2010}).

\bibitem{Zipkes2010}
\bibinfo{author}{Zipkes, C.}, \bibinfo{author}{Palzer, S.},
  \bibinfo{author}{Ratschbacher, L.}, \bibinfo{author}{Sias, C.} \&
  \bibinfo{author}{K$\mathrm{\ddot{o}}$hl, M.}
\newblock \bibinfo{title}{Cold heteronuclear atom-ion collisions}.
\newblock \emph{\bibinfo{journal}{Phys. Rev. Lett.}}
  \textbf{\bibinfo{volume}{105}}, \bibinfo{pages}{133201}
  (\bibinfo{year}{2010}).

\bibitem{Chang2013}
\bibinfo{author}{Chang, Y.-P.} \emph{et~al.}
\newblock \bibinfo{title}{Specific chemical reactivities of spatially separated
  3-aminophenol conformers with cold $\mathrm{Ca}^+$ ions}.
\newblock \emph{\bibinfo{journal}{Science}} \textbf{\bibinfo{volume}{342}},
  \bibinfo{pages}{98--101} (\bibinfo{year}{2013}).

\bibitem{Puri2017}
\bibinfo{author}{Puri, P.} \emph{et~al.}
\newblock \bibinfo{title}{Synthesis of mixed hypermetallic oxide
  $\mathrm{BaOCa}^+$ from laser-cooled reagents in an atom-ion hybrid trap}.
\newblock \emph{\bibinfo{journal}{Science}} \textbf{\bibinfo{volume}{357}},
  \bibinfo{pages}{1370--1375} (\bibinfo{year}{2017}).

\bibitem{Stancil1996}
\bibinfo{author}{Stancil, P.~C.} \& \bibinfo{author}{Zygelman, B.}
\newblock \bibinfo{title}{Radiative charge transfer in collisions of
  $\mathrm{Li}$ with $\mathrm{H}^+$}.
\newblock \emph{\bibinfo{journal}{Astrophys. J.}}
  \textbf{\bibinfo{volume}{472}}, \bibinfo{pages}{102} (\bibinfo{year}{1996}).

\bibitem{Smith1992}
\bibinfo{author}{Smith, D.}
\newblock \bibinfo{title}{{The ion chemistry of interstellar clouds}}.
\newblock \emph{\bibinfo{journal}{Chem. Rev.}} \textbf{\bibinfo{volume}{92}},
  \bibinfo{pages}{1473--1485} (\bibinfo{year}{1992}).

\bibitem{Reddy2010}
\bibinfo{author}{Reddy, V.~S.}, \bibinfo{author}{Ghanta, S.} \&
  \bibinfo{author}{Mahapatra, S.}
\newblock \bibinfo{title}{First principles quantum dynamical investigation
  provides evidence for the role of polycyclic aromatic hydrocarbon radical
  cations in interstellar physics}.
\newblock \emph{\bibinfo{journal}{Phys. Rev. Lett.}}
  \textbf{\bibinfo{volume}{104}}, \bibinfo{pages}{111102}
  (\bibinfo{year}{2010}).

\bibitem{Calvin2018}
\bibinfo{author}{Calvin, A.~T.} \& \bibinfo{author}{Brown, K.~R.}
\newblock \bibinfo{title}{Spectroscopy of molecular ions in coulomb crystals}.
\newblock \emph{\bibinfo{journal}{J. Phys. Chem. Lett.}}
  \textbf{\bibinfo{volume}{9}}, \bibinfo{pages}{5797--5804}
  (\bibinfo{year}{2018}).

\bibitem{Shi2013}
\bibinfo{author}{Shi, M.}, \bibinfo{author}{Herskind, P.~F.},
  \bibinfo{author}{Drewsen, M.} \& \bibinfo{author}{Chuang, I.~L.}
\newblock \bibinfo{title}{Microwave quantum logic spectroscopy and control of
  molecular ions}.
\newblock \emph{\bibinfo{journal}{N. J. Phys.}} \textbf{\bibinfo{volume}{15}},
  \bibinfo{pages}{113019} (\bibinfo{year}{2013}).

\bibitem{Wolf2016}
\bibinfo{author}{Wolf, F.} \emph{et~al.}
\newblock \bibinfo{title}{{Non-destructive state detection for quantum logic
  spectroscopy of molecular ions}}.
\newblock \emph{\bibinfo{journal}{Nature}} \textbf{\bibinfo{volume}{530}},
  \bibinfo{pages}{457} (\bibinfo{year}{2016}).

\bibitem{Rellergert2012}
\bibinfo{author}{{Rellergert}, W.~G.} \emph{et~al.}
\newblock \bibinfo{title}{Evidence for sympathetic vibrational cooling of
  translationally cold molecules}.
\newblock \emph{\bibinfo{journal}{Nature}} \textbf{\bibinfo{volume}{495}},
  \bibinfo{pages}{490--494} (\bibinfo{year}{2012}).

\bibitem{Hudson2016}
\bibinfo{author}{Hudson, E.~R.}
\newblock \bibinfo{title}{Sympathetic cooling of molecular ions with ultracold
  atoms}.
\newblock \emph{\bibinfo{journal}{EPJ Tech. Instrum.}}
  \textbf{\bibinfo{volume}{3}}, \bibinfo{pages}{8} (\bibinfo{year}{2016}).

\bibitem{Hauser2015}
\bibinfo{author}{Hauser, D.} \emph{et~al.}
\newblock \bibinfo{title}{{Rotational state-changing cold collisions of
  hydroxyl ions with helium}}.
\newblock \emph{\bibinfo{journal}{Nat. Phys.}} \textbf{\bibinfo{volume}{11}},
  \bibinfo{pages}{467} (\bibinfo{year}{2015}).

\bibitem{Hudson2018}
\bibinfo{author}{{Hudson}, E.~R.} \& \bibinfo{author}{{Campbell}, W.~C.}
\newblock \bibinfo{title}{{Dipolar quantum logic for freely-rotating trapped
  molecular ions}} (\bibinfo{year}{2018}).
\newblock \bibinfo{note}{{Preprint at https://arxiv.org/abs/1806.09659}}.

\bibitem{Mulin2015}
\bibinfo{author}{Mulin, D.} \emph{et~al.}
\newblock \bibinfo{title}{$\mathrm{H}$/$\mathrm{D}$ exchange in reactions of
  $\mathrm{OH}^-$ with $\mathrm{D}_2$ and of $\mathrm{OD}^-$ with
  $\mathrm{H}_2$ at low temperatures}.
\newblock \emph{\bibinfo{journal}{Phys. Chem. Chem. Phys.}}
  \textbf{\bibinfo{volume}{17}}, \bibinfo{pages}{8732--8739}
  (\bibinfo{year}{2015}).

\bibitem{Allmendinger2016}
\bibinfo{author}{Allmendinger, P.} \emph{et~al.}
\newblock \bibinfo{title}{New method to study ion--molecule reactions at low
  temperatures and application to the reaction}.
\newblock \emph{\bibinfo{journal}{ChemPhysChem}} \textbf{\bibinfo{volume}{17}},
  \bibinfo{pages}{3596--3608}.

\bibitem{Hawley1991}
\bibinfo{author}{Hawley, M.} \& \bibinfo{author}{Smith, M.~A.}
\newblock \bibinfo{title}{Gas phase collisional quenching of $\mathrm{NO}^+$
  (v=1) ions below 5 $\mathrm{K}$}.
\newblock \emph{\bibinfo{journal}{The Journal of Chemical Physics}}
  \textbf{\bibinfo{volume}{95}}, \bibinfo{pages}{8662--8664}
  (\bibinfo{year}{1991}).

\bibitem{Julienne1989}
\bibinfo{author}{Julienne, P.~S.} \& \bibinfo{author}{Mies, F.~H.}
\newblock \bibinfo{title}{Collisions of ultracold trapped atoms}.
\newblock \emph{\bibinfo{journal}{J. Opt. Soc. Am. B}}
  \textbf{\bibinfo{volume}{6}}, \bibinfo{pages}{2257--2269}
  (\bibinfo{year}{1989}).

\bibitem{Gallagher1989}
\bibinfo{author}{Gallagher, A.} \& \bibinfo{author}{Pritchard, D.~E.}
\newblock \bibinfo{title}{Exoergic collisions of cold
  $\mathrm{Na}^{*}$-$\mathrm{Na}$}.
\newblock \emph{\bibinfo{journal}{Phys. Rev. Lett.}}
  \textbf{\bibinfo{volume}{63}}, \bibinfo{pages}{957--960}
  (\bibinfo{year}{1989}).

\bibitem{Weiner1999}
\bibinfo{author}{Weiner, J.}, \bibinfo{author}{Bagnato, V.~S.},
  \bibinfo{author}{Zilio, S.} \& \bibinfo{author}{Julienne, P.~S.}
\newblock \bibinfo{title}{Experiments and theory in cold and ultracold
  collisions}.
\newblock \emph{\bibinfo{journal}{Rev. Mod. Phys.}}
  \textbf{\bibinfo{volume}{71}}, \bibinfo{pages}{1--85} (\bibinfo{year}{1999}).

\bibitem{Gould1998}
\bibinfo{author}{Gensemer, S.~D.} \& \bibinfo{author}{Gould, P.~L.}
\newblock \bibinfo{title}{Ultracold collisions observed in real time}.
\newblock \emph{\bibinfo{journal}{Phys. Rev. Lett.}}
  \textbf{\bibinfo{volume}{80}}, \bibinfo{pages}{936--939}
  (\bibinfo{year}{1998}).

\bibitem{Schowalter2016}
\bibinfo{author}{{Schowalter}, S.~J.} \emph{et~al.}
\newblock \bibinfo{title}{Blue-sky bifurcation of ion energies and the limits
  of neutral-gas sympathetic cooling of trapped ions}.
\newblock \emph{\bibinfo{journal}{Nat. Comm.}} \textbf{\bibinfo{volume}{7}},
  \bibinfo{pages}{12448} (\bibinfo{year}{2016}).

\bibitem{Puri2018}
\bibinfo{author}{Puri, P.}, \bibinfo{author}{Mills, M.}, \bibinfo{author}{West,
  E.~P.}, \bibinfo{author}{Schneider, C.} \& \bibinfo{author}{Hudson, E.~R.}
\newblock \bibinfo{title}{High-resolution collision energy control through ion
  position modulation in atom-ion hybrid systems}.
\newblock \emph{\bibinfo{journal}{Rev. Sci. Instrum.}}
  \textbf{\bibinfo{volume}{89}}, \bibinfo{pages}{083112}
  (\bibinfo{year}{2018}).

\bibitem{Grier2009}
\bibinfo{author}{Grier, A.~T.}, \bibinfo{author}{Cetina, M.},
  \bibinfo{author}{Oru\ifmmode \check{c}\else
  \v{c}\fi{}evi\ifmmode~\acute{c}\else \'{c}\fi{}, F.} \&
  \bibinfo{author}{Vuleti\ifmmode~\acute{c}\else \'{c}\fi{}, V.}
\newblock \bibinfo{title}{Observation of cold collisions between trapped ions
  and trapped atoms}.
\newblock \emph{\bibinfo{journal}{Phys. Rev. Lett.}}
  \textbf{\bibinfo{volume}{102}}, \bibinfo{pages}{223201}
  (\bibinfo{year}{2009}).

\bibitem{Haze2013}
\bibinfo{author}{Haze, S.}, \bibinfo{author}{Hata, S.},
  \bibinfo{author}{Fujinaga, M.} \& \bibinfo{author}{Mukaiyama, T.}
\newblock \bibinfo{title}{Observation of elastic collisions between lithium
  atoms and calcium ions}.
\newblock \emph{\bibinfo{journal}{Phys. Rev. A}} \textbf{\bibinfo{volume}{87}}
  (\bibinfo{year}{2013}).

\bibitem{Chen2014}
\bibinfo{author}{Chen, K.}, \bibinfo{author}{Sullivan, S.~T.} \&
  \bibinfo{author}{Hudson, E.~R.}
\newblock \bibinfo{title}{Neutral gas sympathetic cooling of an ion in a paul
  trap}.
\newblock \emph{\bibinfo{journal}{Phys. Rev. Lett.}}
  \textbf{\bibinfo{volume}{112}}, \bibinfo{pages}{143009}
  (\bibinfo{year}{2014}).

\bibitem{Rouse2017}
\bibinfo{author}{Rouse, I.} \& \bibinfo{author}{Willitsch, S.}
\newblock \bibinfo{title}{Superstatistical energy distributions of an ion in an
  ultracold buffer gas}.
\newblock \emph{\bibinfo{journal}{Phys. Rev. Lett.}}
  \textbf{\bibinfo{volume}{118}}, \bibinfo{pages}{143401}
  (\bibinfo{year}{2017}).

\bibitem{Pechukas1966}
\bibinfo{author}{Pechukas, P.}, \bibinfo{author}{Light, J.~C.} \&
  \bibinfo{author}{Rankin, C.}
\newblock \bibinfo{title}{Statistical theory of chemical kinetics : Application
  to neutral atom--molecule reactions}.
\newblock \emph{\bibinfo{journal}{J. Chem. Phys.}}
  \textbf{\bibinfo{volume}{44}}, \bibinfo{pages}{794--805}
  (\bibinfo{year}{1966}).

\bibitem{Rice1927}
\bibinfo{author}{Rice, O.~K.} \& \bibinfo{author}{Ramsperger, H.~C.}
\newblock \bibinfo{title}{Theories of unimolecular gas reactions at low
  pressures}.
\newblock \emph{\bibinfo{journal}{J. Am. Chem. Soc.}}
  \textbf{\bibinfo{volume}{49}}, \bibinfo{pages}{1617--1629}
  (\bibinfo{year}{1927}).

\bibitem{Dagdigian1977}
\bibinfo{author}{Dagdigian, P.~J.}
\newblock \bibinfo{title}{Dependence of collision complex lifetime on product
  internal state: Laser fluorescence detection of the $\mathrm{Ca}$ +
  $\mathrm{NaCl}$ crossed beam reaction}.
\newblock \emph{\bibinfo{journal}{Chem. Phys.}} \textbf{\bibinfo{volume}{21}},
  \bibinfo{pages}{453 -- 466} (\bibinfo{year}{1977}).

\bibitem{Frisch2009}
\bibinfo{author}{Frisch, M.~J.} \emph{et~al.}
\newblock \bibinfo{title}{{Gaussian 09, Revision B.01}} (\bibinfo{year}{2009}).

\bibitem{Werner2011}
\bibinfo{author}{Werner, H.-J.}, \bibinfo{author}{Knowles, P.~J.},
  \bibinfo{author}{Knizia, G.}, \bibinfo{author}{Manby, F.~R.} \&
  \bibinfo{author}{Sch$\mathrm{\ddot{u}}$tz, M.}
\newblock \bibinfo{title}{Molpro: a general-purpose quantum chemistry program
  package}.
\newblock \emph{\bibinfo{journal}{WIREs Comput. Mol. Sci.}}
  \textbf{\bibinfo{volume}{2}}, \bibinfo{pages}{242--253}
  (\bibinfo{year}{2012}).

\bibitem{Staanum2010}
\bibinfo{author}{Staanum, P.~F.}, \bibinfo{author}{H{\o}jbjerre, K.},
  \bibinfo{author}{Skyt, P.~S.}, \bibinfo{author}{Hansen, A.~K.} \&
  \bibinfo{author}{Drewsen, M.}
\newblock \bibinfo{title}{Rotational laser cooling of vibrationally and
  translationally cold molecular ions}.
\newblock \emph{\bibinfo{journal}{Nat. Phys.}} \textbf{\bibinfo{volume}{6}}
  (\bibinfo{year}{2010}).

\bibitem{Rugango2015}
\bibinfo{author}{Rugango, R.} \emph{et~al.}
\newblock \bibinfo{title}{Sympathetic cooling of molecular ion motion to the
  ground state}.
\newblock \emph{\bibinfo{journal}{N. J. Phys.}} \textbf{\bibinfo{volume}{17}},
  \bibinfo{pages}{035009} (\bibinfo{year}{2015}).

\bibitem{Wagner1972}
\bibinfo{author}{Wagner, A.~F.} \& \bibinfo{author}{Truhlar, D.~G.}
\newblock \bibinfo{title}{Comment on enhancement of the reaction cross section
  of $\mathrm{He} +\mathrm{H}_2^+ \rightarrow \mathrm{He}\mathrm{H}^+ +
  \mathrm{H}$ by vibrational excitation of $\mathrm{H}_2^+$ and the treatment
  of nuclear spin by the statistical phase‐space theory}.
\newblock \emph{\bibinfo{journal}{J. Chem. Phys.}}
  \textbf{\bibinfo{volume}{57}}, \bibinfo{pages}{4063--4064}
  (\bibinfo{year}{1972}).

\bibitem{Pechukas1965}
\bibinfo{author}{Pechukas, P.} \& \bibinfo{author}{Light, J.~C.}
\newblock \bibinfo{title}{On detailed balancing and statistical theories of
  chemical kinetics}.
\newblock \emph{\bibinfo{journal}{J. Chem. Phys.}}
  \textbf{\bibinfo{volume}{42}}, \bibinfo{pages}{3281--3291}
  (\bibinfo{year}{1965}).

\bibitem{Mills2017}
\bibinfo{author}{Mills, M.} \emph{et~al.}
\newblock \bibinfo{title}{Efficient repumping of a ca magneto-optical trap}.
\newblock \emph{\bibinfo{journal}{Phys. Rev. A}} \textbf{\bibinfo{volume}{96}},
  \bibinfo{pages}{033402} (\bibinfo{year}{2017}).

\end{thebibliography}

\appendix
	
\section{Reaction rate measurements} \label{RPC}

In general, the BaCl$^+$ sample may react with several Ca electronic levels that are simultaneously populated, each of which may have differing spatial density distributions and chemical reactivities. Consequently, the decay of BaCl$^+$ from the LQT can be modeled using a simple rate equation model as
\begin{equation}
\frac{d}{dt}N_{BaCl^+} = -\Gamma_T N_{BaCl^+}
\end{equation}
where $\Gamma_T$ is the total reaction rate and is calculated as
\begin{equation} \label{rateconstant}
\Gamma_{T} = \sum\limits_{i}{\int_{}^{} \hat\rho_{ion}(\vec{r}) \eta_{i}(\vec{r}) k_{i}(E_{col}(\vec{r})) d^{3}r}
\end{equation}
where the sum is carried over all populated Ca levels, $\eta_{i}(\vec{r})$ is the peak-normalized atom number density of the $i^{th}$ Ca electronic state at position $\vec{r}$, $\hat\rho_{ion}(\vec{r})$ is the integral-normalized ion-number density a evaluated at position $\vec{r}$ (time-dependent if shuttled), and $k_{i} (E_{col}(\vec{r}))$ is the reaction rate constant of the $i^{th}$ electronic state at the spatially dependent collision energy, $E_{col}(\vec{r})$. Note that the energy of the ions when the sample is held stationary is determined primarily by the position-dependent micromotion energy. However, when shuttled, the ion sample also possesses kinetic energy from the time-dependent shuttle trajectory, and $\Gamma_{T}$ is determined by taking the time-average of equation~(\ref{rateconstant}) over the experimental duty cycle. After the total reaction rate has been determined, the individual rate constants for the various Ca electronic states can be inferred. 

\begin{figure}
\begin{center}
\includegraphics[width=70 mm]{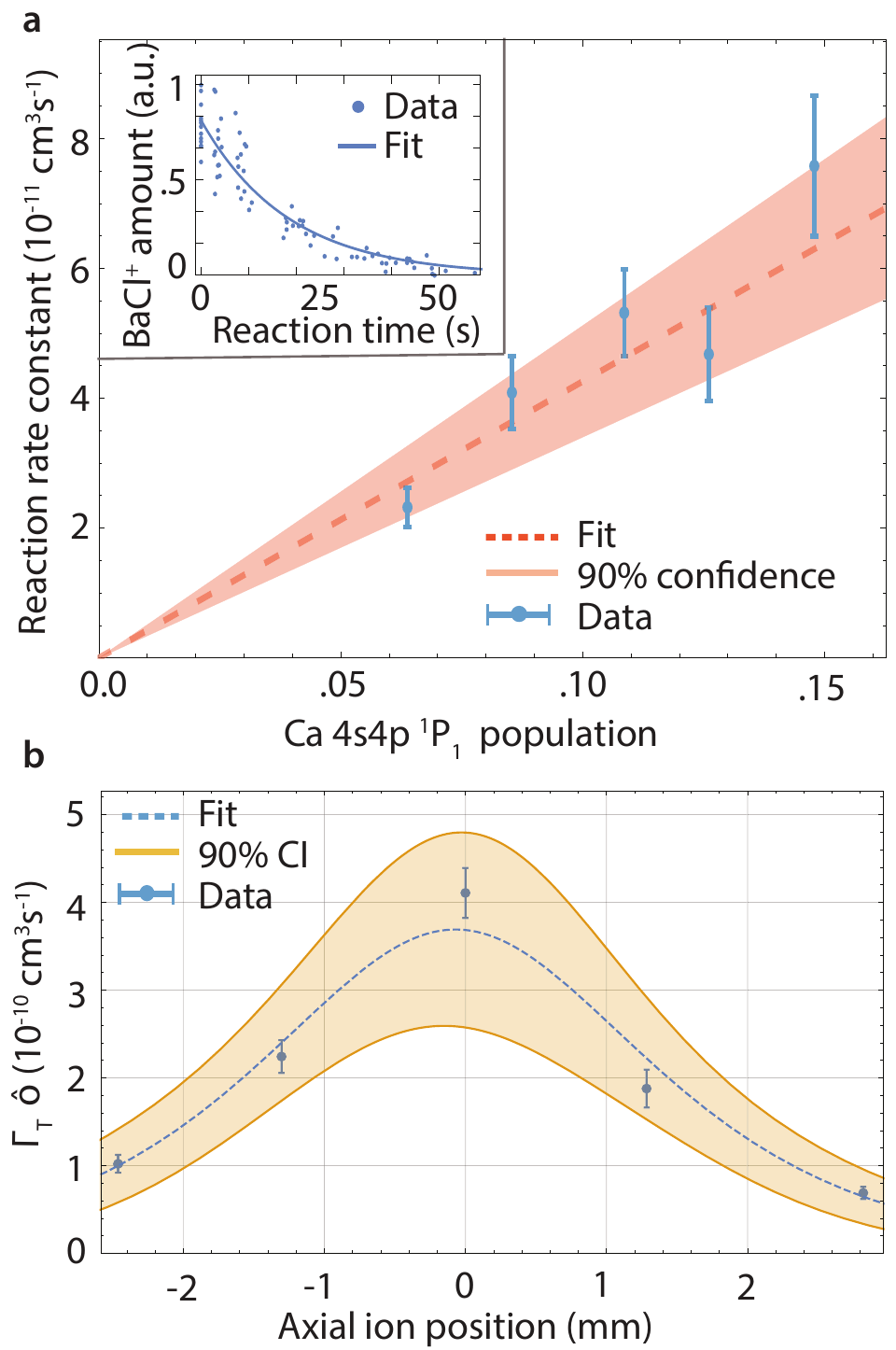}
\end{center}
\caption{\textbf{Reaction rate characterization} \\
\textbf{(a)} Observed reaction rate constant dependence on Ca 4s4p $^1$P$_1$ population. The inset displays a typical reactant decay curve used to extract the rates, with the reactant amount determined by integrating over ToF-MS spectra. \textbf{(b)} The measured Ca 4s4p $^3$P$_2$ reaction rate constant, $\Gamma_{\mathrm{T}}$, multiplied by the geometric atom-ion overlap factor, $\hat{\mathrm{O}}$, obtained at different spatial offsets between the ion sample and the center of a magnetic trap of pure triplet atoms. The corresponding fit curve (dashed line) along with its $90\%$ confidence interval (CI) (yellow band) are displayed as well. The functional form of the fit curve (equation~(\ref{mag})) allows for approximate estimation of the magnetic trap density profile. For both plots, each data point consists of approximately 100 measurements, where error bars represent one standard error.}
\label{fig:RPC}
\end{figure}

MOT laser cooling parameters are varied to determine the relationship between the reaction rate and atomic population in both the 4s4p $^1$P$_1$  and 4s4p $^3$P$_J$ states (Appendix Fig.~\ref{fig:RPC}a). For the 4s4p $^1$P$_1$ state, the intensity of the cooling beam is manipulated to vary the singlet state population while the reaction kinetics data is extracted, yielding a reaction rate constant of $6(1.5) \times 10^{-10}$ cm$^3$s$^{-1}$ for the 4s4p $^1$P$_1$ state (obtained at a collision energy of $\approx 10$~K). 

Typically, the MOT triplet steady-state population is $\approx100\times$ smaller than the singlet population, making it difficult to distinguish between reactions originating from the triplet state and those originating from the singlet. In order to extract reliable 4s4p $^3$P$_J$ reaction rate constants, in a manner similar to Ref.~\cite{Puri2017}, BaCl$^+$ ions are overlapped with a magnetic trap of 4s4p $^3$P$_2$ atoms lost from the MOT laser cooling cycle~\cite{Puri2017}. The magnetic trap spatial profile needed for atom number density calculations, and thus rate constant extraction, is obtained by monitoring how the reaction rate varies as a function of displacement from the trap center (Appendix Fig.~\ref{fig:RPC}b). Using this method at a collision energy of $\approx 10$~K, an absolute reaction rate constant of $2(1) \times 10^{-9}$~cm$^3$s$^{-1}$ is extracted for the 4s4p $^3$P$_J$ state, a rate roughly a factor of four larger than that of its 4s4p $^1$P$_1$ counterpart. 

In order to calculate the spatial overlap of the magnetic trap and the ions, the center of the magnetic trap is first identified by observing a Ba$^+$ laser-cooling fluorescence dip due to coherent-population-trapping (CPT) at the magnetic field zero of the system. Subsequently, a bias current is applied to the anti-Helmholtz coils utilized for constructing the MOT trapping potential. The bias current shifts the magnetic field zero, and thus the center of the magnetic trap, by a fixed distance, $z_0$, away from the ions (typically chosen to be $\approx500~\mu$m) in the axial dimension, ensuring that CPT effects do not complicate reaction rate data acquisition. The radial distance of the ions from the center of the magnetic trap is then tuned by adjusting the endcap voltages of the LQT, and reaction rates are measured at various atom-ion spatial offsets. At each position in the LQT, the ions experience a local magnetic field density, $\rho_{M}(r)$, of 

\begin{equation} \label{mag}
\rho_{M}(r) =\rho_{M_0} e^{\frac{-\sqrt{r^2 + 4z_0^2}}{\omega_B}}
\end{equation}
where $\rho_{M_0}$ is the peak density of the magnetic trap, $r$ is the radial position of the ions with respect to the magnetic trap center, and $\omega_B$ is the spatial length scale of the magnetic trap, determined as $\omega_B=(0.98 \pm 0.11)$~mm through fits to the above-mentioned reaction rate data (error expressed at the 68$\%$ confidence interval). For clarity, we emphasize that the axial and radial dimensions referenced here refer to the conventional dimensions of the anti-Helmholtz coil geometry and not the LQT trapping potential.    
 
While other excited states are occupied during laser cooling, such as the 4s4p $^1$D$_2$ state, these states have substantially smaller electronic state populations than the 4s4p $^1$P$_1$  and 4s4p $^3$P$_J$ levels and are not expected to contribute substantially to the observed reaction rates.

\section{Phase space theory calculation of branching fractions} \label{phasespace}

Under the assumption of strong coupling, all electronic, orbital, and angular momenta are expected to mix. Thus, while each individual angular momentum is not a conserved quantity throughout the reaction, the total angular momentum, $K$, along with its cylindrical axis projection, $K_z$, are conserved~\cite{Wagner1972}, with $K$ being the magnitude of the vector sum
\begin{equation}
\mathbf{K} = \mathbf{\Bell} + \mathbf{N} + \mathbf{L} + \mathbf{S}
\end{equation}
where $\mathbf{\Bell}$, $\mathbf{N}$, $\mathbf{L}$, and $\mathbf{S}$ are the vectors for the orbital, rotational, electronic orbital, and electronic spin angular momenta of the reaction complex. For the reactions studied in this work, $|\mathbf{L}|$ and $|\mathbf{S}|$ are $\leq$1, while generally $|\mathbf{\Bell}|$ and $|\mathbf{N}|$ can often exceed $\approx$ $10$, meaning the former can be neglected in the following calculation for simplicity. 

\begin{figure}
\begin{center}
\includegraphics[width=6cm]{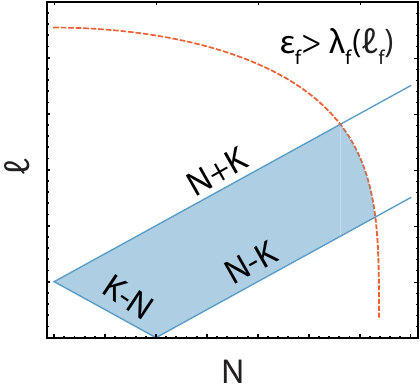}
\end{center}
\caption{\textbf{Phase space diagram for branching fraction calculation} \\
A phase space diagram showing the range of rotational (N) and orbital ($\ell$) angular momentum product states accessible at a given reactant total angular momentum (K). The shaded region of the curve denotes the final states that both obey angular momentum conservation and possess enough product kinetic energy, $\epsilon_f$, to clear the product state centrifugal energy barrier, $\lambda_f(\ell_f)$, and dissociate from the three body reaction complex into the final product atom and diatomic molecule.}
\label{fig:BR}
\end{figure}

In both the reactant ($i$) and product ($f$) states, $K$ is bounded as 
\begin{equation} \label{rest1}
|\ell_{i(f)}-N_{i(f)}| \leq K \leq |\ell_{i(f)}+N_{i(f)}|
\end{equation}

Additionally the final product must possess enough kinetic energy to escape the angular momentum barrier of the exit channel, permitting the three-body complex to dissociate into its molecular and atomic constituents. The later restriction is satisfied by enforcing
\begin{equation}\label{rest2}
\epsilon_f \geq \lambda_f(\ell)
\end{equation}
where $\epsilon_f$ is the final kinetic energy of a given product state and $\lambda_f(\ell_f)$ is the height of the centrifugal barrier in each product exit channel at orbital angular momentum $\ell_f$. $\epsilon_f$ can be calculated as
\begin{equation}
\begin{split}
\epsilon_f (\epsilon_i,v_i,N_i,v_f,&N_f,Q_{i,f}) =\\ 
 &\epsilon_i + E_{int} (v_i,N_i) - E_{int} (v_f,N_f) + Q_{i,f} 
\end{split}
\end{equation}
where $\epsilon_i$ is the initial collision energy of the reaction complex, $Q_{i,f}$ is the exothermicity of reaction for exit channel $f$, and
\begin{equation}
\begin{split}
E_{int} (v_{i(f)},N_{i(f)}) &= \omega_{i(f)} (v_{i(f)}+1/2) \\
&+ 2 \beta_{i(f)} N_{i(f)} (N_{i(f)}+1) \\
&-\omega_{i(f)}\chi_{i(f)} (v_{i(f)}+1/2)^2
\end{split}
\end{equation}
is the internal energy in the reactant (product) state associated with the $v_{i(f)}$ and $N_{i(f)}$ vibrational and rotational level, respectively, which is characterized by the spectroscopic constants $\omega_{i(f)}$, $\beta_{i(f)}$, and $\omega_{i(f)}\chi_{i(f)}$.

Equations (\ref{rest1}) and (\ref{rest2}) restrict the number of $\ell_f$ and $N_f$ states accessible to each exit channel (see Appendix Fig.~\ref{fig:BR}) at a given total angular momentum, $K$, and exit-channel kinetic energy, $\epsilon_f$. By counting the number of states accessible to each possible exit channel at given $K$ and $\epsilon_f$, one can define the probability of accessing each exit channel as
\begin{equation}
\begin{split}
&P_f(K,\epsilon_f (\epsilon_i,v_i,N_i,v_f,N_f,Q_{i,f})) \\
&= \frac{ \sum_{v_f,N_f}d_f\overline{n_f}(K,\epsilon_f (\epsilon_i,v_i,N_i,v_f,N_f,Q_{i,f}))}{\sum_b \sum_{v_b,N_b}d_b\overline{n_b}(K,\epsilon_b (\epsilon_i,v_i,N_i,v_b,N_b,Q_{i,b}))}
\end{split}
\end{equation}
where $\overline{n_b}(K,\epsilon_b (\epsilon_i,v_i,N_i,v_b,N_b,Q_{i,b}))$ is the the total number of states accessible for exit channel $b$ at a given $K$ and $\epsilon_b$, and $d_b$ is a degeneracy factor that accounts for the spin multiplicity of each product state. We note that $\overline{n_b}(K,\epsilon_b (\epsilon_i,v_i,N_i,v_b,N_b,Q_{i,b}))$ is proportional to the area bounded by curves presented in Appendix Fig.~\ref{fig:BR}.

Therefore, again following Ref.~\cite{Pechukas1965}, the total reaction cross-section for a given exit channel, $f$, given an initial reactant rotational quantum number, $N_i$ (assuming the reactant is in the ground vibrational state), and summed over all accessible product rotational and vibrational states is given as 

\begin{equation}
\begin{split}
\sigma_f (N_i,\epsilon_i) &= \sum_{\ell_i=0}^{\ell_{\rm{max}}(\epsilon_i)} \frac{2(\ell_i+1) \pi \hbar^2}{2 \mu_f \epsilon_i} P_f(\ell_i) \\
&= \sum_{\ell_i=0}^{\ell_{\rm{max}}(\epsilon_i)} \frac{\pi \hbar^2}{2 \mu_f \epsilon_i (2N_i+1)} \sum_{|\ell_i-N_i| \leq \atop K \leq |\ell_i+N_i|} (2K+1) \\
& \times P_f(K,\epsilon_f (\epsilon_i,v_i,N_i,v_f,N_f,Q_{i,f})) 
\end{split}
\end{equation}
where $\mu_f$ is the reduced mass of the product complex.

Lastly, to calculate the final branching fractions, we must average each cross-section across the rotational temperature of the sample as 
\begin{equation}
\begin{split}
\overline{\sigma_f} & = \sum_{N_i}\frac{1}{Z}(2 N_i+1) \\ 
&\times e^{-2 \beta_i N_i(N_i+1)/(k_B T)} \sigma_f (N_i,\epsilon_i)
\end{split}
\end{equation}
where $T\approx 2\epsilon_i/k_B$ is the effective rotational temperature of the initial reactant molecular ion and $Z$ is the rotational partition function. 

Finally, after the relevant cross-sections have been computed, the branching fraction into each exit channel, $\gamma_f$, is given by
\begin{equation} \label{finalBR}
\gamma_f = \frac{\overline{\sigma_f}}{\sum_{f} \overline{\sigma_f}}
\end{equation} 
Equation~(\ref{finalBR}) is applied to the Ca$^*$+BaCl$^+$ system, and the results are compared directly to experimental branching fractions in the main text (Fig. 2c). In addition to the product branching fractions, equation~(\ref{finalBR}) is also applied to estimate the percentage of collisions that occur and result in excited states of [BaCl$^+$ + Ca] that ultimately radiatively decay back into the [BaCl$^+$ + Ca] ground state. This factor is included as $\chi_S$ in equation (6) of the main text and adjusts our rate constant calculation to account for experimentally indistinguishable events where inelastic collisions occur but no new molecular products are formed (see Appendix~\ref{MCT}). Errors in the calculated branching ratios can be primarily attributed to uncertainties in the exit channel exothermicities, nonergodicity in the system, and uncertainties in the molecular constants used in the state counting process.  

\section{Modified capture theory calculation} \label{MCT}

A semi-classical model for the reaction blockading effect was constructed that considers the effect of each participating partial wave on the total reaction cross-section by solving an Einstein rate equation model for the Ca atom in the presence of a laser field. As the atom approaches the ion along its reactive trajectory, the energy spacing between the ground and excited state changes from its separation limit value due to the differing long-range forms of the respective potentials. Thus, any laser fields that are resonant with the atoms in the separation limit become detuned as the reaction trajectory proceeds, resulting in a reduced excited-state population at short-range as compared the steady-state value.

To account for this effect and to calculate the proper population fractions at short-range, the model assigns the optical field a time-dependent laser detuning, $\Delta(t) = [V_{ex}(r(t),\ell)-V_{gs}(r(t),\ell)]/\hbar$, where $V_{ex}(r(t),\ell)$ is the excited-state molecular potential, $V_{gs}(r(t),\ell)$ is the ground-state molecular potential, $r(t)$ is the time-dependent internuclear-separation distance, and $\hbar$ is the reduced Planck's constant.

Two levels are considered in the rate equation model, the ground Ca $^1$S$_0$ state (N$_{0,0}$) and the excited Ca $^1$P$_1(m_J=0)$ state (N$_{1,0}$). The system of equations is constructed as follows

\begin{equation}
\begin{split}
\frac{dN_{0,0}}{dt} &= B_{10}(t)(N_{1,0}-N_{0,0})\\
\frac{dN_{1,0}}{dt} &= A_{10} N_{1,0} + B_{10}(t)(N_{0,0}-N_{1,0}) \\
\end{split}
\end{equation}

where $N_{J,m_J}$ refers to population in the $(J,m_J)$ state and $A_{10}$ is the Einstein-A coefficient between the two levels. $B_{10}(t)$ is the time-dependent Einstein-B coefficient between the two levels that incorporates $\Delta(t)$ and is given as
\begin{equation}
B_{10}(t) = \Gamma_{10} \frac{\pi^2 c^3}{\hbar \omega_{10}^3}\frac{I}{2 \pi c}\frac{\Gamma_{10}}{(\delta-\Delta(t))^2+(\Gamma_{10}/2)^2} 
\end{equation}
where $\Gamma_{10}$ is the transition linewidth, $\omega_{10}$ is the transition energy, and $\delta$ is the separation-limit laser frequency detuning from the atomic transition.

$r(t)$ is determined by solving Newton's classical law of motion in the collision frame, $\mu \ddot{r}=\frac{d}{dr} V_{av}(r)$. To estimate the force the system experiences as it transitions between the ground and excited state potentials, we use an averaged potential, $V_{av}(r)$, that is weighted by the ground-state and excited-state populations as

\begin{equation}
V_{av}(r) = 
\begin{cases}
      \rho_{pp} V_{ex}(r,\ell) + (1-\rho_{pp}) V_{gs}(r,\ell) & r>r_c \\
			& \\
      V_{gs}(r,\ell) & r \leq r_c
   \end{cases}
\end{equation}
where $\rho_{pp}$ is the steady-state excited-state fraction of the Ca MOT~\cite{Mills2017} and $r_c$ is defined in equation $(6)$ of the manuscript ($\rho_{pp}\approx0.2$ under standard laser-cooling conditions). In the model, the averaged potential is approximated as transitioning to a fully ground-state potential at $r_c$. This approximation is made since at $r_c$, the addressing laser field is detuned by one linewidth from resonance, meaning the atoms primarily reside in their ground state potential. An abrupt shift in potential was chosen over continuous updating of the weightings of the averaged potential for computational convenience.

The initial conditions for the equation of motion are chosen as follows: $	r(0) = 3000$~nm and $\dot{r}(t)=(2E_{col}/\mu)^{1/2}$. The computation was repeated with $r(0)$ being varied across $1000$~nm, confirming that choice of initial condition did not affect the trajectory result within a $<1\%$ level. Both $V_{ex}(r(t),\ell)$ and $V_{gs}(r(t),\ell)$ contain the orbital-angular-momentum-dependent centrifugal barrier term, $\hbar^2 \ell(\ell+1)/(2 \mu r^2)$, where $\ell$ is the orbital angular momentum of the system, ensuring that each partial wave has a unique collision trajectory.   

The reaction blockade factor for each partial wave, $P_{\ell}(E_{col},m_J)$, can be determined by dividing the excited-state fraction predicted by the rate equation model at short-range by its long-range value, $\rho_{pp}$. The separation distance at which the chemical event occurs is $r_s$, and therefore the excited-state fraction for the $m_J=0$ projection state at short-range is given as $N_{1,0}(t_c)$ where $t_c$ is defined as $r(t_c)=r_c$. Consequently, $P_{\ell}(E_{col},m_J)=N_{1,m_J}(t_c) / \rho_{pp}$, and the total cross-section, including the blockading effect, can be calculated as
\begin{equation}\label{eqn:MCT}
\begin{split}
\sigma(&E_{col}) \approx\\
 &\frac{\pi \hbar^2}{2 \mu E_{col}} \sum_{m_J} \frac{ \eta_{m_J} \chi_S}{2J+1} \sum_{\ell=0}^{\ell_{\rm{max}}(E_{col},m_J)} (2\ell+1) P_{\ell} (E_{col},m_J) 
\end{split}
\end{equation}
where all quantities are defined as they are in equation (6) of the main text. The collision energy of the system sets the initial velocity of the collision complex and has a significant effect on how likely the Ca atom is to reach short-range before spontaneously decaying. We note that exclusion of dark states, such as the Ca 3d4s $^1$D$_2$ state, in the model is warranted since over the collision energy range studied, the atom-ion complex accelerates to short-range in $<10$ atomic lifetimes, and the branching fraction into dark levels from the excited Ca singlet state is $1:10^5$. 

The sum over all $\ell$ and $m_J$ states in equation~(\ref{eqn:MCT}) is carried out for both the triplet and singlet manifold, with the results plotted alongside the data in Fig. 2a in the main text. We also note for all molecular potentials with repulsive long-range forms, such as the $(J,m_J)$=$(1,\pm1)$ projection states, the reactants repel one another and thus do not reach short-range, resulting in $P_{\ell}(E_{col},\pm1)=0$. Lastly, for further explanation of the $\eta_{m_J}$ factors in the above equation as well as a more thorough discussion of cases where $P_{\ell}(E_{col},m_J)=0$, refer to the proceeding section.

\section{Cross-sections and electronic structure calculations} \label{theory}

As depicted in Fig.~3a,e (main text), the effect of the quadrupole moment on the long-range curves is non-trivial, leading to barriers that reduce reaction rates for some channels or more attractive curves that increase the reaction rates for others.

Consequently, the long-range curves have a strong effect on the modified capture theory presented in Appendix~\ref{MCT}. The theoretical cross-sections are dependent on the factor $P_{\ell} (E_{col},m_J)$ (see equation~(\ref{eqn:MCT})), which is only non-zero if the long-range potential for a particular $m_J$ projection-state is attractive. Therefore, if a $J$ manifold has repulsive $m_J$ projection states, its average cross-section, and also its reaction rate, will be reduced accordingly.

The long-range curves plotted in Fig. 3a,e of the main text can be used to determine these reduction factors for both the singlet and triplet manifolds. Let $\sigma^{S/T}_{J,|m_J|}(E)$ correspond to the energy-dependent cross-section for the $(J,|m_J|)$ projection state within either the singlet (S) or triplet (T) manifold. For the singlet manifold corresponding to BaCl$^+$ + Ca ($^1$P$_1$), we find that both $m_J = \pm 1$ projection states corresponding to the $(J,|m_J|)=(1,1)$ curve are repulsive (Fig. 3a, main text). Therefore, at low energy, both $P_{\ell} (E_{col},|m_J|)$ and $\sigma^{S}_{J,|m_J|}(E_{col})$ are approximately zero for these states. The total cross-section for the singlet manifold is given by the sum over all projection states, as described in equation~(\ref{eqn:MCT}), leading to a total singlet cross-section of
\begin{equation}
\begin{split}
   \sigma^{S}(E_{col}) &= \sum_{m_J} \sigma^{S}_{J,|m_J|}(E_{col})\\
	&= \frac{2}{3} \sigma^S_{1,1}(E_{col}) + \frac{1}{3} \sigma^S_{1,0}(E_{col}) \\
	&\approx \frac{1}{3} \sigma^S_{1,0}(E_{col})
	\end{split}
\end{equation}

Similarly, in the triplet case, since only the $J=2$ of Ca ($^3$P$_J$) was excited when reaction rate data was experimentally acquired, we find that
\begin{equation}
\begin{split}
   \sigma^{T}(E_{col}) &= \sum_{m_J} \sigma^{T}_{J,|m_J|}(E_{col})\\
	&= \frac{2}{5} \sigma^T_{2,2}(E_{col}) + \frac{2}{5} \sigma^T_{2,1}(E_{col}) + \frac{1}{5} \sigma^T_{2,0}(E_{col}) \\
	&\approx \frac{2}{5} \sigma^T_{2,1}(E_{col})+\frac{1}{5} \sigma^T_{2,0}(E_{col})
	\end{split}
\end{equation}
again due to the energetic barrier for ($J,|m_J|)$=(2,2) projection states (Fig. 3e, main text). Note that for all cases studied in this work, the potentials for both $(J,\pm m_J)$ are degenerate. 

For a complete reaction rate calculation, the results from the long-range potential energy surfaces (PESs) must be matched the shorter-range PESs; consequently, electronic structure calculations are performed to calculate the latter. The reaction surface for each excited-state reagent is computed using coupled cluster theory including single and double excitations (CCSD). For the electronically excited states that correlate to Ca$^*$ + BaCl$^+$, with Ca (4s4p $^1$P$_1$) and Ca (4s4p $^3$P$_2$), reaction surfaces are computed using equation-of-motion coupled cluster theory including single and double excitations (EOM-CCSD). The reaction surface calculations are performed on a grid of $133$ points at various angles and distances of approach using the triple-zeta correlation consistent basis sets (cc-pwCVTZ-PP on calcium and barium and cc-pVTZ on chlorine). The core electrons in calcium and barium are replaced by effective core potentials. The electronic structure calculations are performed using the Gaussian $09$~\cite{Frisch2009} and Molpro $2012$~\cite{Werner2011} program packages.

The potential energy surfaces (PESs) shown in Fig.~3 (main text) for the three excited singlet and triplet symmetries, $1A'$, $2A'$ and $A''$, are produced by interpolating the $133$ points computed on a grid ranging from $4$ to $10$ $\AA$ and angles between $0$ and $180^{\circ}$ (i.e. from Ca approaching BaCl$^+$ from Cl to Ca approaching BaCl$^+$ from Ba, and angles between these two linear approaches corresponding to $0$ and $180^{\circ}$, respectively). The dark plane in each panel indicates the asymptotic energy of BaCl$^+$ + Ca$^*$, 3.08 eV for singlet and 1.88 eV for triplet.

For the singlet case, each of the three long-range curves obtained by considering BaCl$^+$
as a point charge will be mapped roughly equally onto the three shorter-range PESs, as the 
BaCl$^+$ molecular axis is randomly oriented along the $x$, $y$, or $z$ axis in the laboratory frame
defining the $m_J$ long-range projections. We account for this by assigning a probability of 1/3 to each
of the $(J,m_J)$ curves to correspond to the $1A'$, $A''$, or $2A'$ PES at shorter-range, and write
the total probability of reaching the short-range reaction region, $\eta_{S}$, as 
\begin{equation}
    \eta_{S} = \frac{1}{3} \eta_{1A'} + \frac{1}{3} \eta_{A''} + \frac{1}{3} \eta_{2A'} \;,
\end{equation}
where $\eta_{S}$ is the probability of reaching the short-range reaction region for the singlet PES. As shown
in Fig.~3 (main text), there is a barrier preventing Ca$^*$ to reach that region for both PES
corresponding to $A''$ and $2A'$, while no such barrier exists for $1A'$, i.e. 
$\eta_{A''} = \eta_{2A'} = 0$ and $\eta_{1A'} = 1$, leading to $\eta_{S}$= 1/3.

In the case of the triplet manifold, all three short-range PESs corresponding to $1A'$, $A''$ and 
$2A'$ allow Ca$^*$ to reach the reaction region, {\it i.e.} $\eta_{1A'}=P_{A''}=\eta_{2A'}=1$. In this case, any long-range curve allowing Ca$^*$ to reach short-range will also permit the system to evolve in the reaction region, resulting in
\begin{equation}
    \eta_{T}= 1
\end{equation}
where $\eta_{T}$ is the probability of reaching the short-range reaction region for the triplet PES.

After $\eta_{T}$ and $\eta_{S}$ have been determined, they can be inserted into equation~(\ref{eqn:MCT}) to yield the total cross-section and, consequently, also the reaction rate constant. These curves are plotted alongside the data in Fig. 2a of the main text.

\end{document}